\documentclass[aps, prb, twocolumn, groupedaddress, showkeys]{revtex4}
\usepackage[dvips]{color}   
\usepackage{xcolor}
\usepackage{amsfonts}
\usepackage{amsmath}
\usepackage{amssymb}
\usepackage{graphicx} 
\usepackage{mathrsfs}   
\usepackage{bbm} 
\usepackage{bm}
\usepackage{multirow}
\usepackage[sort&compress]{natbib}
\usepackage{dcolumn}%

\begin{document}

\newcommand{\dd}{d}
\newcommand{\pd}{\partial}
\newcommand{\myU}{\mathcal{U}}
\newcommand{\myr}{q}
\newcommand{\Urho}{U_{\rho}}
\newcommand{\myalpha}{\alpha_*}
\newcommand{\bd}[1]{\mathbf{#1}}
\newcommand{\Eq}[1]{Eq.~(\ref{#1})}
\newcommand{\Eqn}[1]{Eq.~(\ref{#1})}
\newcommand{\Eqns}[1]{Eqns.~(\ref{#1})}
\newcommand{\Figref}[1]{Fig.~\ref{#1}}
\newtheorem{theorem}{Theorem}
\newcommand{\me}{\textrm{m}_{\textrm{e}}}
\newcommand{\sgn}{\textrm{sign}}
\newcommand*{\bfrac}[2]{\genfrac{\lbrace}{\rbrace}{0pt}{}{#1}{#2}}

\title{Geodesy on surfaces of revolution: A wormhole application}

\author{Lorenzo Gallerani Resca}          
\email{resca@cua.edu}
\homepage{http://physics.cua.edu/people/faculty/homepage.cfm}
\thanks{Corresponding Author.}

\affiliation{Department of Physics and Vitreous State Laboratory, 
The Catholic University of America,
Washington, DC 20064} 

\author{Nicholas A. Mecholsky}
\email{nmech@vsl.cua.edu}

\affiliation{Department of Physics and Vitreous State Laboratory, 
The Catholic University of America,
Washington, DC 20064}

\date{\today}  

\begin{abstract}

We outline a general procedure to derive first-order differential equations obeyed by geodesic orbits over two-dimensional (2D) surfaces of revolution immersed or embedded in ordinary three-dimensional (3D) Euclidean space. We illustrate that procedure with an application to a wormhole model introduced by Morris and Thorne (MT), which provides a prototypical case of a `splittable space-time' geometry. We obtain analytic solutions for geodesic orbits expressed in terms of elliptic integrals and functions, which are qualitatively similar to, but even more fundamental than, those that we previously reported for Flamm's paraboloid of Schwarzschild geometry. Two kinds of geodesics correspondingly emerge. Regular geodesics have turning points larger than the `throat' radius. Thus, they remain confined to one half of the MT wormhole. Singular geodesics funnel through the throat and connect both halves of the MT wormhole, perhaps providing a possibility of `rapid inter-stellar travel.' We provide numerical illustrations of both kinds of geodesic orbits on the MT wormhole.

\end{abstract}




\maketitle   

\section{Introduction}  
The study of geodesy has been instrumental to the development of civilization in science and engineering since time immemorial. In the western world, not to mention others, it provided foundations to mathematics, astronomy and geography, at least since Eratosthenes and the Greek golden era, through Gauss and the developers of non-Euclidean geometries, till space-time geometrodynamics in and of the cosmos as begun by Einstein, just to name some of the great masters and recall major advances.     

In this grand scheme, it is still exciting to provide even some relatively minor contribution, but valuable nonetheless, as we do in this paper. At an elementary technical level, we describe a general procedure to derive first-order differential equations obeyed by geodesic orbits over two-dimensional (2D) surfaces of revolution immersed or embedded in ordinary three-dimensional (3D) Euclidean space. We illustrate our mathematical procedure with some applications to axially symmetric wormholes of notable interest in general relativity (GR). 

\section{Geodesy on surfaces of revolution}\label{revolution-surfaces}
      
Let us first establish integrable geodesic orbit equations on regular two-dimensional surfaces $S$ of revolution. Points $\bar{p}$ on $S$ are most readily parameterized by transforming from Cartesian coordinates, $(x,y,z)$, to cylindrical coordinates, $(r,\phi,z)$, so that                   
\begin{align}\label{point}
\bar{p}  =  & ( r \cos \phi, r \sin \phi, z) .
\end{align}
In the $x-z$ plane, a \textit{profile curve} is represented by a smooth function, $r=r(z)$, which is then rigidly rotated by any angle $\phi$ between $0$ and $2\pi$ around the $z-$axis, thus providing a surface of revolution $S$ with azimuthal symmetry. Typically, one may also locally invert a profile curve as $z=z(r)$ and transform correspondingly differential elements or equations.           

In this latter perspective, we may represent line elements in terms of metric tensor components on $S$ as 
\begin{equation}\label{line-element}
(dl)^2  =  g_{ij} dx^i dx^j =  g_{rr} (dr)^2 + g_{\phi \phi} (d\phi)^2 ,
\end{equation}
where summation is implied over repeated dummy indices as in $dx^i = (dr, d \phi)$ and 
\begin{equation}\label{r-metric-component}
g_{rr} = 1 + \bigg(\frac{dz}{dr}\bigg)^2 ,
\end{equation}
\begin{equation}\label{phi-metric-component}
g_{\phi \phi} = r^2 .
\end{equation}

A \textit{curve} $\gamma(\lambda)$ lying on $S$ is parameterized in terms of two $x^i = x^i(\lambda)$ coordinates in \Eq{line-element}. \textit{Tangent vectors} to $\gamma(\lambda)$ on $S$ thus have two $V^i=\frac{d x^i}{d \lambda}$ components. \textit{Geodesic curves} for \textit{contravariant} components generally satisfy \textit{second-order} differential equations of the form 
\begin{equation}\label{geodesic-Christoffel}
  \frac{d V^i}{d \lambda} + \Gamma^i_{jk} V^j V^k = 0 ,
\end{equation}
where $\Gamma^i_{jk}$ represent well-known Christoffel symbols.\cite{Pressley,SchutzGM} 

Introducing \textit{covariant} components, 
\begin{equation}\label{covariant-components}
V_i  =  g_{ij} V^j ,
\end{equation}
equivalent geodesic equations 
\begin{equation}\label{geodesic-covariant}
  \frac{d V_k}{d \lambda}= \frac{1}{2} \bigg( \frac{\partial g_{ij}}{\partial x^k} \bigg) V^i V^j 
\end{equation}
can be generally derived.\cite{Schutz2Ed} Both forms of geodesic equations inherently define classes of \textit{affine parameters} $\lambda$ linearly related to each other. 

When there are symmetries in $S$, \Eq{geodesic-covariant} is most suitable to provide corresponding conserved quantities. For surfaces of revolution $S$ with azimuthal symmetry, we characteristically obtain
\begin{equation}\label{angular-momentum}
 V_{\phi} = g_{\phi \phi} V^{\phi} = r^2 \bigg(\frac{d \phi}{d \lambda}\bigg) = L = \mathrm{const}.
\end{equation}

In a mechanical analog, \Eq{angular-momentum} provides a \textit{`first integral'} of the motion that represents conservation of an angular momentum component $L$ along the symmetry axis of revolution of $S$. In elementary differential geometry, the same result is obtained more laboriously and it is known as Clairaut's relation or theorem.\cite{Pressley}

\textit{Parallel transport} by geodesic curves of their tangent vectors further requires conservation of the vector's \textit{norm} or length. For geodesics on $S$ we thus obtain 
\begin{equation}\label{spatial-norm}
  V_i V^i= g_{ij} V^i V^j = g_{rr} \big(\frac{dr}{d \lambda}\big)^2 + \frac{L^2}{r^2} = C^2 > 0 ,
\end{equation}
where \Eq{angular-momentum} has been introduced in \Eq{line-element}. In a mechanical analog, the positive constant $C^2$ relates to conservation of energy and provides yet another central \textit{`first integral'} of the motion.

Thus, without any need to further integrate either geodesic \Eq{geodesic-Christoffel} or \Eq{geodesic-covariant}, we are able to obtain from \Eq{spatial-norm} a \textit{first-order} decoupled differential equation
\begin{equation}\label{radial-geodesic}
  \bigg(\frac{dr}{d\lambda}\bigg)^2
    = \big(\frac{1}{g_{rr}}\big) \big(C^2- \frac{L^2}{r^2}\big),
\end{equation}
which can be directly integrated by separation of variables. 

Surprisingly, these latter elementary but critical steps for establishing and solving geodesic equations on surfaces of revolution \textit{in general} are missing in elementary textbooks on differential geometry, which thus remain stuck with the much more complicated \Eq{geodesic-Christoffel}. For example, on p. 227, Pressley\cite{Pressley} states that ``geodesic equations for a surface of revolution cannot usually be solved explicitly.'' In fact, \Eq{radial-geodesic} allows us to do precisely that. We shall not report in this paper corresponding results for classical geodesic problems on ellipsoids, hyperboloids, or quadric surfaces of revolution, but our readers may derive any such results for themselves by direct application of our simpler technique.

Further progress can be made by considering \textit{turning points}, defined as having 
\begin{equation}\label{turning-point}
  \bigg(\frac{dr}{d\lambda}\bigg)_{r_t}  = 0 .  
\end{equation}
Thus we obtain
\begin{equation}\label{turning-point-value}
  {r_t}^2 = \frac{L^2}{C^2} ,
\end{equation}
provided that $1/{g_{rr}}$ does not vanish for a greater $r$-value. 

In terms of $r_t$, by taking the ratio between the two first integrals in \Eq{radial-geodesic} and \Eq{angular-momentum} for $L \ne 0$, we can eliminate the presence of any affine parameter and obtain a \textit{geodesic orbit} equation
\begin{equation}\label{geodesic-orbit}
  \bigg( \frac{dr}{d\phi} \bigg)^2
    = \bigg( \frac{r^2}{g_{rr}} \bigg) \bigg( \frac{r^2}{r_{t}^2} - 1\bigg) .
\end{equation}
This represents our \textit{central result}. It can be directly solved for any regular surface $S$ of revolution by quadrature or numerical integration after separation of variables.

It is also possible to show that
\begin{equation}\label{Clairaut}
r \sin{\psi} = r_0 \sin{\psi_0} = L/C = \pm{r_t} = \mathrm{const} 
\end{equation} 
along a geodesic, which represents precisely Clairaut's relation.\cite{Pressley} Here $\psi$ represents the angle that a parallel-transported vector $\bar{V}$ forms with the local meridian passing through $(r, \phi)$ on $S$. Thus, parallel transport from an initial vector $\bar{V}_0$ tangent to $S$ yields a \textit{geodesic orbit} equation
\begin{equation}\label{geodesic-orbit-initial-vector}
  \bigg( \frac{dr}{d\phi} \bigg)^2
    = \bigg( \frac{r^2}{g_{rr}} \bigg) \bigg( \frac{r^2}{[r_{0} \sin{ \psi_0}]^2}  - 1\bigg) .
\end{equation}
Evidently, \Eq{geodesic-orbit} and \Eq{geodesic-orbit-initial-vector} are equivalent on account of \Eq{Clairaut}, although differently parametrized. One may use one geodesic orbit equation or the other, depending on one's choice of initial or search conditions.        

Furthermore, \Eq{turning-point} means that $r_t$ is an extremal point. Correspondingly, \Eq{Clairaut} implies that at $r = r_t$ we must have $\psi = \psi_t = \pm \pi/2$, meaning that a geodesic orbit must become extremally tangent to a local parallel at a turning point. Overall, however, parallels are not geodetic, except for an equatorial parallel, where $z = z(r)$ is also extremal. 

A local inversion $z=z(r)$ has been typically assumed in \Eq{r-metric-component} and in the resulting geodesic orbit \Eq{geodesic-orbit}. Returning to the original single-valued $r=r(z)$ profile curve, the equivalent geodesic orbit equation is          
\begin{equation}\label{geodesic-orbit-equivalent}
  \bigg( \frac{dz}{d\phi} \bigg)^2
    = \bigg( \frac{[r(z)]^2}{g_{zz}} \bigg) \bigg( \frac{[r(z)]^2}{[r(z_{t})]^2} - 1\bigg) ,
\end{equation}
where   
\begin{equation}\label{z-metric-component}
g_{zz} = 1 + \bigg(\frac{dr}{dz}\bigg)^2 ,
\end{equation}
\begin{equation}\label{z-turning-point-value}
 [r(z_{t})]^2 = {r_t}^2 = \frac{L^2}{C^2} .
\end{equation}

\section{Geodesy on wormholes in general relativity}\label{wormholes}

Using this technique, we have solved analytically and numerically geodesic orbit equations in general relativity for Schwarzschild geometry, e.g., on Flamm's paraboloid of revolution.\cite{Resca, Rafael} That is equivalent to a so-called `Schwarzschild wormhole' or a corresponding `splittable space-time' metric.\cite{MTW, Rindler, Wald, Ohanian, D'Inverno, Hartle, Hobson, Frolov, Carroll, Poisson, PricePrimer, Price, PricePRD} The latter sets $g_{tt} = - 1$, which prevents geodesics from penetrating the event horizon.\cite{PricePRD}   

Interestingly, Eqs. (11) and (26) of Ref.~\onlinecite{Rafael} confirm that a particle placed at rest on any point of Flamm's paraboloid remains at rest therein,\cite{Price} rather than eventually falling toward the hole. Namely, having set $C = L = 0$ initially, the particle maintains zero velocity and zero acceleration equivalents indefinitely. That further dispels a common analogy of relativistic space curvature as the distorsion of an elastic sheet caused by a massive ball weighing on a spandex,\cite{Resca, Price, Middleton, Janis} as often portrayed even on covers of outstanding textbooks.\cite{D'Inverno, Carroll, Poisson}         

Extending Schwarzschild geometry with Kruskal-Szekeres (KS) coordinates demonstrates that this metric is not fully static and the spatial equatorial representation as a Flamm's paraboloid applies exactly only at a $v=0$ pseudo-time.\cite{MTW, Rindler, Wald, Ohanian, D'Inverno, Hartle, Carroll, Hobson, Frolov} Evolving $v$ between space-like hypersurfaces with $v=-1$ and $v=+1$, a variously called Einstein-Rosen\cite{EinsteinRosen} `bridge' or Schwarzschild `throat' or KS `wormhole' first opens and widens and then closes again. However, that happens so fast that even null radial world-lines can hardly traverse such wormhole.     

Consequently, an entire field of inquiry has developed regarding questions as to whether `short-cuts' through space-time or related `time travel machines' may develop through similar structures. In this context, we may only quote a few references that may serve as entry points to a diverse historical, technical, philosophical and science-fiction literature.\cite{Wheeler, Thorne, Visser, Raine, Lockwood} 

To illustrate our geodetic technique beyond our previous applications to Flamm's paraboloid,\cite{Resca, Rafael} we consider here a prototypical most instructive wormhole model introduced by Morris and Thorne (MT).\cite{MorrisThorne} In order to form a `throat' theoretically suitable for rapid interstellar travel by human beings, MT began by constructing a well-behaved wormhole geometry that has no event horizon nor any curvature singularity. Their geometry is also asymptotically flat, spherically symmetric, non-rotating and fully static, thus avoiding any possibility of closed time-like world-lines with their associated anti-causal paradoxes.\cite{Visser, Lockwood} The MT wormhole is also a prototype of a `splittable space-time,' built of a parallel stack of identical spatial geometries.\cite{PricePRD} From a so constructed geometry, MT calculate Riemann and Ricci tensors and then the stress-energy tensor that correspondingly derives from Einstein's field equations. As one may have expected, regions of negative energy density occur, violating classical energy conditions. However, the possibility that quantum field effects may allow microscopically some sort of those violations to occur remains open. Whatever the case, the MT wormhole model has acquired great pedagogical value and widespread consideration.

Thus MT arrive to the splittable space-time metric
\begin{align}\label{STwormholesphericalmetric}
ds^2  = & -(cdt)^2 + (dR)^2 + ( b^2 + R^2) \bigg( (d \theta)^2 + \sin^2 \theta (d \phi)^2 \bigg)
\end{align}
in $(R, \theta, \phi)$ spherical coordinates, where $0 \le R < \infty$. Given the spherical symmetry, geodesics must be equatorial, thus $\theta = \frac{\pi}{2} = \mathrm{const}$ may be assumed for those. 

Transforming to cylindrical coordinates $(r,\phi,z)$ with the restriction that
\begin{align}\label{restriction}
r^2  = b^2 + R^2 ,
\end{align}
we obtain by isometry that
\begin{align}\label{STwormholecylindricalmetric}
ds^2   = & -(cdt)^2 + \bigg(1 - \frac{b^2}{r^2} \bigg)^{-1} (dr)^2 + r^2 (d \phi)^2 .
\end{align}
As in Schwarzschild geometry, the $r$ curvature coordinate generates a coordinate singularity at $r = b > 0$ in \Eq{STwormholecylindricalmetric}, but  that has no penetrable horizon on account of \Eq{restriction}.

With our procedure, it is straightforward to find either \textit{time-like} or \textit{null} geodesic equations and orbits for \Eq{STwormholecylindricalmetric} and express them in terms of unique turning points. Remarkably, as is the case for the `splittable space-time' reduction of Schwarzschild metric,\cite{Resca, Rafael} all those geodesics formally coincide with \textit{space-like} geodesics on the `a-temporal' MT wormhole,\cite{PricePRD} having a positive-definite metric
\begin{align}\label{MTwormholecylindricalmetric}
(dl)^2   = \bigg(1 - \frac{b^2}{r^2} \bigg)^{-1} (dr)^2 + r^2 (d \phi)^2 
\end{align}
at any $t = \mathrm{const}$ slicing.

Now \Eq{MTwormholecylindricalmetric} has the form of \Eq{line-element} and our general results apply. By embedding the 2D `a-temporal' MT wormhole in 3D Euclidean space, it is easy to find that its profile curve is
\begin{align}\label{MTwormhole}
r(z) = b \cosh (z/b) .
\end{align}
That follows from integration and inversion of \Eq{r-metric-component}, using \Eq{MTwormholecylindricalmetric} to express $g_{rr}$. For an alternative but equivalent derivation, see for instance Sec.~7.7 and Figs.~(7.4) and (7.5), pp.~148-152, of Hartle's book,\cite{Hartle} keeping in mind that Hartle denotes our $R$ as his `$r$' and our $r$ as his `$\rho$'.

Thus, according to \Eq{radial-geodesic}, we immediately obtain the equatorial geodesic equation
\begin{equation}\label{MTradial-geodesic}
  \bigg(\frac{dr}{d\lambda}\bigg)^2
    = \big( 1 - \frac{b^2}{r^2} \big) \big(C^2- \frac{L^2}{r^2}\big)
\end{equation}
for the MT wormhole. That has a turning point given by \Eq{turning-point-value} if that $r_t^2$ exceeds $b^2$. Otherwise, the turning point occurs at $b = r_t'$, although we may still formally maintain the notation and relation
\begin{equation}\label{singular-turning-point-value}
  b^2 \equiv ({r_t'})^2 > \frac{L^2}{C^2} \equiv {r_t}^2 .    
\end{equation}

According to \Eq{geodesic-orbit}, we thus obtain the \textit{geodesic orbit} equation 
\begin{equation}\label{MTgeodesic-orbit}
  \bigg( \frac{dr}{d\phi} \bigg)^2
    = ( r^2 - b^2 ) \bigg( \frac{r^2}{r_{t}^2} - 1\bigg) 
\end{equation}
for the MT wormhole. Now that can be directly integrated by separation of variables as 
\begin{equation}\label{spatialMTsolutionintegral}
\phi(r) =  \int \frac{ \dd r}{\sqrt{( r^2 - b^2 ) ( \frac{r^2}{r_{t}^2} - 1 )}} ,
\end{equation} 
generating analytic solutions expressed in terms of elliptic integrals and functions. We report here some of those expressions, which are qualitatively similar to, but even more fundamental than, those that we have already reported for Flamm's paraboloid.\cite{Rafael}

For regular or non-singular orbits, having $r_t>b$ and $n$-labeled, we obtain   
\begin{equation}
   \phi_{\textrm{n}}(r) =  \textrm{K}\left[ \frac{b^2}{r_t^2}\right]-\textrm{F}\left[\sin
   ^{-1}\left(\frac{r_t}{r}\right) \bigg| \frac{b^2}{r_t^2}\right] ,
\end{equation}
where $\textrm{F}\left[ \phi | m \right]$ is the incomplete elliptic integral of the first kind,
\begin{equation}
\textrm{F}\left[ \phi | m \right] = \int_{0}^{\phi} \left(1 - m \sin^2 \theta\right)^{-1/2} \, \dd \theta,
\end{equation}
for $-\pi/2 < \phi < \pi/2$. Extensions beyond this range of $\phi$ may be made using transformations of the argument as
\begin{equation}
\textrm{F}\left[n \pi \pm \phi| m \right] = 2 n \textrm{K}[m] \pm \textrm{F}\left[\phi | m \right],
\end{equation}
where $\textrm{K}\left[ m \right] = \textrm{F}\left[\pi/2 | m \right]$ is the complete elliptic integral of the first kind.\cite{Library}

This gives upon inversion,      
\begin{equation}
   r(\phi) = r_t \, \textrm{dc} \left(\phi \left| \frac{b^2}{r_t^2} \right.\right) ,
\end{equation}
where dc denotes the Jacobi $dc$ elliptic function, and the angle $\phi$ is taken within the range $(-\phi_{n \infty}, \phi_{n \infty})$, where
\begin{align}\label{eqn:phiinfinity}
\phi_{n \infty} &= \lim_{r \to \infty} \phi_{\textrm{n}}(r) = \textrm{K} \left[ \frac{b^2}{r_t^2} \right].
\end{align}       

For singular orbits, having $r_t < b$ and $s$-labeled, we obtain 
\begin{equation}
   \phi_{\textrm{s}} (r) = \frac{1}{b} r_t \left(\textrm{K}\left[\frac{r_t^2}{b^2}\right]-\textrm{F}\left[\sin
   ^{-1}\left(\frac{b}{r}\right) \bigg| \frac{r_t^2}{b^2}\right] \right) .
\end{equation}

This gives upon inversion,
\begin{equation}
   r_{\textrm{s}} (\phi) = b \, \text{dc}\left(\frac{b \, \phi }{r_t} \bigg| \frac{r_t^2}{b^2} \right) ,
\end{equation}
where dc denotes again the Jacobi $dc$ elliptic function and the angle $\phi$ is taken within the range $(-\phi_{s \infty}, \phi_{s \infty})$, where
\begin{align}\label{eqn:phiinfinity}
\phi_{s \infty} &= \lim_{r \to \infty} \phi_{\textrm{s}}(r) = \frac{r_t}{b} \textrm{K}\left[\frac{r_t^2}{b^2}\right] .
\end{align}

Rather than using the $r$ coordinate, we may equivalently derive geodesic orbit equations in terms of either $R$ or $z$ coordinates. Those have the advantage of removing the coordinate singularity at $r = b$, as we have similarly demonstrated on Flamm's paraboloid.\cite{Rafael} Geometrically, azimuthal revolution of the profile curve \Eq{MTwormhole} generates both halves of the MT wormhole joined at its `throat.' Thus, essentially the same geodesic structure and completeness that we already found on Flamm's paraboloid appears on the MT wormhole. Two kinds of geodesics correspondingly emerge. If ${r_t}^2 = \frac{L^2}{C^2}$ exceeds $b^2$, \textit{regular} geodesics reach that turning point and remain confined to one half of the MT wormhole. Otherwise, \textit{singular} geodesics (as expressed in terms of the singular curvature coordinate $r$) reach the `throat' or minimal circle at $r = b$ and funnel between both halves of the whole MT wormhole. Both kinds of geodesics may or may not encircle the hole region any number of times, crossing themselves correspondingly. Infinitely many geodesics can possibly be drawn between any two points, but they must be of specific regular or singular types.

\begin{figure}[!hb]      
	\begin{center}
 	\includegraphics[width=8.6cm]{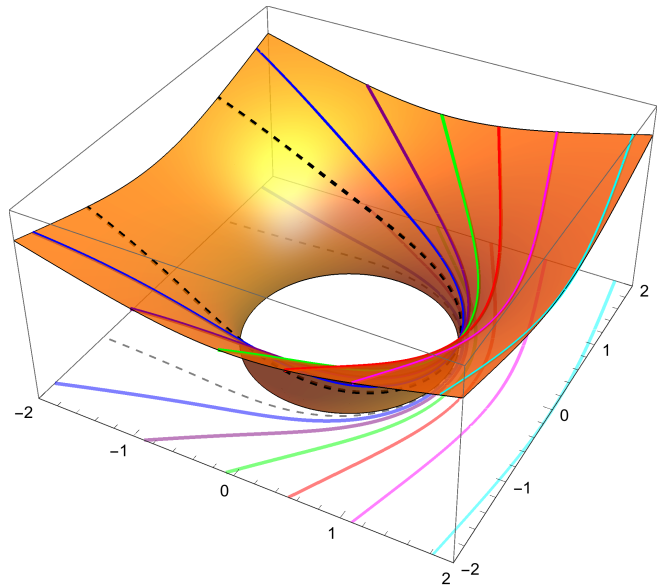}\\
      \caption{\label{fig:MTembeddedgeodesics} Regular geodesics embedded on half of the MT wormhole with turning points at $r_t =$ 2.0, 1.5, 1.25, 1.125, 1.0625, 1.03125, and 1.01581. Here $b = 1$. The critical $2\pi$-encircling geodesic orbit with $r_{t1} = 1.01581 $ is displayed as a dashed line.}  
	\end{center}
\end{figure}

Some regular geodesic orbits embedded on half of the MT wormhole are graphed in \Figref{fig:MTembeddedgeodesics} for turning point values $r_t > b = 1$. Based on azimuthal symmetry, the $x$-axis has been aligned along the direction from $r = b = 1 $ to $r_t$.

\begin{figure}[!hb]
	\begin{center}
	\includegraphics[width=8.6cm]{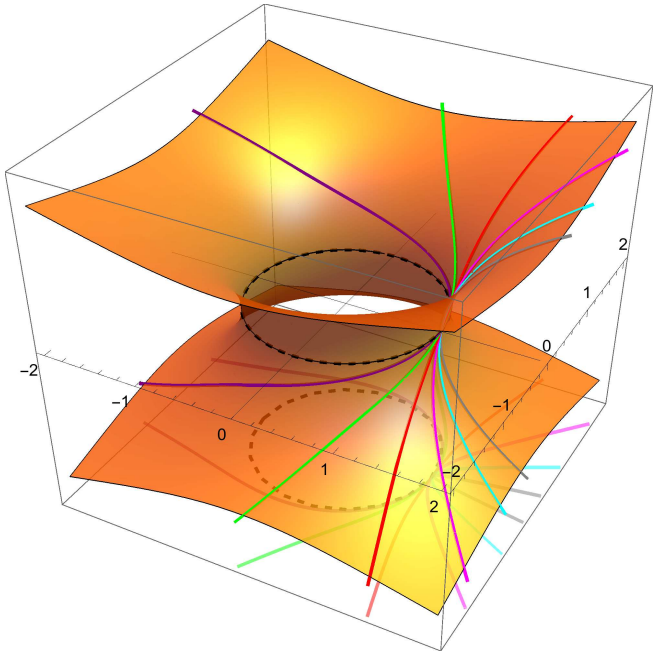}\\
      \caption{\label{fig:MTembeddedgeosb} Singular geodesics with varying $1 > r_t =$ 0.1, 0.3, 0.5, 0.7, 0.9, 0.99 values. The geodesic dashed circle provides the turning point radius $r_t' = b = 1$ common to all singular geodesics. The closer the $r_t$ parameter is to 1, the more a singular geodesic may wind around the hole. In this Figure, only the innermost purple singular geodesic with the greatest $r_t = 0.99$ value fully encircles the hole region once, as clear from its planar projection.}
	\end{center}
\end{figure}

Examples of singular geodesics embedded on the full MT wormhole are shown in \Figref{fig:MTembeddedgeosb}. The dashed circle represents the geodesic orbit separating singular from regular geodesics. Speculatively, `interstellar travel' could take place along singular geodesics, being shortest and fastest along radial geodesics with $L = r_t = 0$.

For the MT wormhole, geodesic results equivalent or complementary to ours have been previously obtained.\cite{PricePRD, MullerAJP, MullerPRD} In particular, Eq.~(12) of Ref.~\onlinecite{MullerPRD}, derived from a Lagrangian formalism,\cite{Rindler} corresponds to our \Eq{radial-geodesic}. However, the development of initial conditions in Ref.~\onlinecite{MullerPRD} differs from our formulation in terms of turning points. Which results are more practical or essential depends to some extent on the objectives of each inquiry.

\section{Conclusions}

We have outlined a general procedure to derive first-order differential equations obeyed by geodesic orbits over two-dimensional (2D) surfaces of revolution immersed or embedded in ordinary three-dimensional (3D) Euclidean space. We have illustrated that procedure with an application to a wormhole model introduced by Morris and Thorne (MT).\cite{MorrisThorne} That provides a prototypical case of a `splittable space-time' geometry,\cite{PricePRD} in which all geodesics formally coincide with \textit{space-like} geodesics on the `a-temporal' MT wormhole, having the same positive-definite metric at any $t = \mathrm{const}$ slicing. We have obtained analytic solutions for geodesic orbits expressed in terms of elliptic integrals and functions, which are qualitatively similar to, but even more fundamental than, those that we previously reported for Flamm's paraboloid of Schwarzschild geometry.\cite{Rafael} Two kinds of geodesics correspondingly emerge. Regular geodesics have turning points larger than the `throat' radius. Thus, they remain confined to one half of the MT wormhole. Singular geodesics funnel through the throat and connect both halves of the MT wormhole, perhaps providing a possibility of `rapid inter-stellar travel.' We have provided numerical illustrations of both kinds of geodesic orbits on the MT wormhole.

\acknowledgments

We acknowledge financial support from the Vitreous State Laboratory at the Catholic University of America.

\bibliographystyle{apsrev4-1}


\end{document}